# Economic dimension of crimes against cultural-historical and archaeological heritage  (EN)

Shteryo Nozharov[1]

The publication is one of the first studies of its kind, devoted to the economic dimension of crimes against cultural and archaeological heritage. Lack of research in this area is largely due to irregular global prevalence vague definition of economic value of the damage these crimes cause to the society at national and global level, to present and future generations. The author uses classical models of Becker and Freeman, by modifying and complementing them with the tools of economics of culture based on the values of non-use. The model tries to determine the opportunity costs of this type of crime in several scenarios and based on this to determine the extent of their limitation at an affordable cost to society and raising public benefits of conservation of World and National Heritage.

### Introduction

Research problem – significance and relevance.

According to the database of National Statistic Institute (NSI) of Bulgaria, the share of tourism of GDP for 2013 is 13.6%, as the share of cultural tourism in the tourist sector is 12% (ITCG, 2014). This means that the revenues of cultural tourism are over 1 billion levs per year in the conditions of post recession recovery of Bulgaria when the economic growth of the country is 0.9% in 2013 according to the database of NSI. In accordance with the National strategic plan for cultural tourism development, adopted by the Ministry of culture, the structure of the cultural tourism includes: cultural heritage tourism, art tourism, creative tourism. Inspite there is no information about the share of direct revenues of cultural heritage tourism in the framework of cultural tourism, probably its share is significant as Bulgaria is famous for its 1300 years history and through its territory they are passing many corridors of cultural tourism because of its neighbourhood with Turkey and Greece, which are two of the leading countries in this field. In addition to the share of GDP, cultural tourism contributes to the balanced regional development of countries, as it creates employment in the regions with high unemployment rate.

---

[1] Shteryo Nozharov is a chief assistant professor, PhD, full faculty member at Department of Economics, UNWE and also a barrister, e-mail: nozharov@unwe.bg ;

*This study was published for the first time in 2015 only in Bulgarian. The current version is its first English translation.*

On the other hand, the cultural tourism potential could not be beneficially used without the protection and exploration of objects of movable and immovabale cultural-historical and archaeological heritage. The potection of objects of cultural-historical and archaeological heritage is firmly linked with the issues, concerning the national identity of the country, as well as with the sustainable development and what we leave for the future generations. For Bulgaria, this issue is of international importance, having in mind its membership in UNESCO, but it also influences Bulgaria's international authority and has economic and geostrategic effects.[i] In view of this, the indirect economic effects of protection of objects of cultural-historical and archaeological heritage could not be neglected.

That is why, the crimes against cultural-historical and archaeological heritage damage the national economy and have many economic dimensions, which could be a subject and an object of scientific research.

Degree of elaboration of the research problem in the scientific literature:

In general, the problems, directly linked to economics of crime in the international literature are new problems in the field of the economic science, developed in the 60s years of the XX$^{th}$ century and here we can highlight the names of some of the Nobel laureates such as Gary Becker (University of Chicago), Richard Posner (University of Chicago), Richard B. Freeman (Harvard University), Richard Rosenfeld (University of Missouri-St. Louis) and others.

However, these research studies are conceptual and examines what the effects of the classical types of crimes, such as crimes against humanity (murders, body injuries), crimes against property (robbery, theft), trafficking and distribution of prohibited and dangerous substances (drugs, weapons, toxic chemicals) are.

There was not found any publication in the field of economic aspects of crimes against cultural-historical and archaeological heritage in the literature review of foreign research studies.

The problems, related to economics of crimes in general are also examined in the Bulgarian scientific literature:
- The Center for the Study of Democracy has written short articles on the topic of economics of crime, which are rather strategically and sociologically oriented than economically oriented (CSD 2002; CSD 2003);
- The Institute for Market Economics presents a short review of the foreign publications in the field of economics of crime (Ganev, 2008);

The only specific study directly linked to the topic of economic aspects of crimes against cultural-historical and archaeological heritage is that of the Center for the Study of Democracy in 2007 in relationship to a criminological analysis of the organized crime in Bulgaria. This study examines not only the drug markets, human trafficking, prostitution, market of stolen cars, but also the antiques market (CSD, 2007). The publication presents short sociological information about the trends in this types of crimes and addresses the roles and crminological motives of the main participants in the antiques market. However, the publication has no economic analysis because they are presented only numerous citations of regulations.

The presented publications do not cover the main focus of the currents research and consequently, it could be assumed that the research is devoted to a relevant problem which is slightly examined in the Bulgarian literature and it is a new problem worldwide

The relevance of the problem in theoretical aspect is related to the examination of the economic aspects which influence and emerge from the crimes against cultural-historical and archaeological heritage.

The relevance of the problem has also a practical aspect, related to:

1. Possible change in the methodology of the forensic archaeological expertise, while considering the total economic value of loss of public welfare, caused by this type of crimes.
2. Introducing an econometric basis for establishing the program budget of the Ministry of Culture in the field of protection of movable and immovable cultural heritage values and possible change in the program budgets of institutions responsible for the pre-trial procedures and investigation of this type of crimes by considering the scope of the total loss of public welfare and counteraction of this crimes;
3. Guidelines for changing the scope of cultural tourism potential and assessing the effects of this aspects over the contribution of tourism in GDP.

Object and subject of the research. Research statement and restrictive conditions.

The object of the research is the *economic dimensions of crimes* and its subject are *crimes against cultural-historical and archaeological heritage.*

Predicate – the research feature is the unused potential of „values of non-use ", used mainly in the environmental economics for determination of the total economic value of crimes and of crimes against cultural-historical and archaeological heritage.

*This study was published for the first time in 2015 only in Bulgarian. The current version is its first English translation.*

Defining the object and subject of the research:

The current research confirms the following working definitions:
- Definition for „*cultural-historical heritage*", given in art.2, p.1 of the Cultural Heritage Law, according to which the objects of cultural-historical heritage are bearers of historical memory, national identity and have scientific and cultural value.
- Definition for „*archaeological heritage*", given in art.146, p.1 of the Cultural Heritage Law, according to which the objects of archaeological heritage are traces of human activity of past epochs, located or found in the earth layers, surface or underwater for which the main source of information are the archaeological excavations;
- Definition for "*crimes against cultural-historical and archaeological heritage*", given in art.208 and from art.277a to art.278b of the Criminal Code of Republic of Bulgaria, which will be studied in depth in the next sections of the current research.

In the current research, it is made a distinction between crimes against cultural-historical heritage and crimes against archaeological heritage due to the following reasons:

*First*, a similar distinction is made in the Bulgarian Criminal Code. In the separate criminal panels, as an object of crimes are determined either archaeological site (art.277, p.1-8; art.278,p.6 of the Criminal Code) or cultural value (art.208, p.4; art.278, p.1-5; art.278a and art.278b of the Criminal Code).

This normative distinction of crimes predefines the statistical and econometric distinction when grouping the data gathered by the judicial bodies and measuring the values – the subject of the research.

*Second*, a similar distinction is made also in the Bulgarian Cultural Heritage Law. In art.47, p.1, archaeological values are determined as a sub-type of the cultural values. There is also a special section, named "Archaeological cultural heritage" in this Law, which highlights the special status of this type of cultural value for the society. In the administrative-penalty provisions of this Law, it is made a distinction between an illegal act, affecting cultural-historical value (art.200a and b, art.203 and etc.) and an illegal act, affecting archaeological value (art.200c, art.218a and etc.).[ii]

This normative distinction of crimes predefines the statistical and econometric distinction when grouping the data gathered by the administrative and penalty bodies and measuring the values – a subject of the research.

Introducing a hypothetic definition of „opportunity costs of crimes against cultural-historical and archaeological heritage "so as to be determined the economic dimension of this type of crimes:

*This study was published for the first time in 2015 only in Bulgarian. The current version is its first English translation.*

The opportunity costs of crimes against cultural-historical and archaeological heritage, considered as the value of the best alternative we have rejected in a situation of choice, is the difference amongst:

- First alternative: not to counteract strongly against these types of crimes (type of counteraction "as before"), which alternative is presented in the third section of the study.
- Second alternative: to enhance the counteraction of these types of crimes and to value the benefits of the protected cultural-historical and archaeological values from criminal encroachments (for example: cultural tourism, clarifying the national history and identity, image of Republic of Bulgaria as one of the few countries which owns most of the cultural values in the world cultural heritage and etc.).
- Third alternative: to counteract at the maximum to this type of crimes, as we need to redirect the budget constraint and human resources of the police, prosecutor's offices and justice in order to protect the cultural-historical values for the future generations in Bulgaria. This fact is important, as the cultural heritage of Bulgaria could be a competitive advantage in the future, because the country has a rich history in an environment where the world culture is put on the equal.

The study was conducted in the following restrictions:
1. There is a lack of sufficiently published statistical data for the criminal justice in Bulgaria.
2. The study is focused on Bulgaria, despite the conclusions made in the final section have a universal character and could be successfully used in the criminal justice practices in other countries.
3. There is a lack of national funding for conducting a representative sociological research for the needs of the survey.

Goal and tasks of the research. Research hypothesis

The main goal of the current study is to be defined and analyzed the components of the total economic value of objects of cultural-historical and archaeological heritage and as a result to be highlighted the guidelines for determining the opportunity costs of crimes against cultural-historical and archaeological heritage.

The research hypothesis is that:
The economic dimension, identified as opportunity costs of crimes against cultural-historical and archaeological heritage, could not be calculated through the classical economic models, based on the direct consumer value without considering the values of non-use, but carrying consumer or social utility and related to the social welfare.

Working hypothesis:
- Economic dimensions of crimes against cultural-historical and archaeological heritage are unstudied field in the economic science;
- The opportunity cost of crimes against cultural-historical and archaeological heritage could not be calculated through the classical economic models, based on the direct consumer value;
- The values of non-use out of the direct consumer value but carrying consumer or social utility and related to the social welfare, could contribute to the measuring of the total economic value of crimes against cultural-historical and archaeological heritage.

The proof of the formulated thesis and working hypothesis and having in mind the restrictions of the study, put the need the following tasks to be fulfilled:
1. To be made a review of the economic theory in the field of economics of crimes and to be analyzed those statements that have in common with the fulfillment of the main goal of the current research.
2. To be developed conceptual guidelines for determination of the total economic value of objects of cultural-historical and archaeological heritage.
3. Based on the determination of the total economic value of objects of cultural-historical and archaeological heritage to be determined the opportunity cost of crimes against cultural-historical and archaeological heritage.

Theoretical and methodological basis of the research. Applied methods.
In methodological aspect, the following approaches are applied - systematic, comparative, and statistical, as the studied economic processes are examined in a continued and close relationship and development one to another.
The theoretical methods of synthesis and analysis are done through their corresponding approaches of induction and deduction, which approaches are related to the application of the abstraction method.
In spite of the definitive character of the induction approach, in the research it is taken into account the general theory in the relevant field but also the possibility to make sense of the facts not only to

*This study was published for the first time in 2015 only in Bulgarian. The current version is its first English translation.*

arrange them in a logical way. In the way, deduction method is used, the author put the accent over the nature of the facts to avoid acquiring stochastic character.

The methods are subordinated directly to the main goal of the study and this fact predefines the logical structure of the research:

Through the method of induction, the general principles of *opportunity costs* and *economic value* as economic categories are highlighted, while the deduction method puts the accent over the specific forms of these principles when the economic dimension through opportunity costs of crimes against cultural-historical and archaeological heritage is determined.

The study is based on a comparative analysis, which includes processes and categories which show themselves on a national scope (for Bulgaria) but the data derived could be successfully used for comparisons, concerning other countries.

Some categories and processes are examined as dynamic ones, which means that the counteraction of crimes is considered as a process, which could be developed or changed over time.

There is separate statistical analysis in the research, which is related to the studying of correlations amongst direct consumer value, value of non-use, consumer, or social utility, related to the social welfare. All these categories are studied to be determined the opportunity costs of crimes against cultural-historical and archaeological heritage.

The structure of the research is done in accordance with the understanding that the opportunity costs of crimes against cultural-historical and archaeological heritage as economic dimension for the present and future generations is much greater than the society's costs for counteraction of these crimes and this fact is necessary to lead to a change in the present models of calculation, assessment and counteraction of these crimes.

Information provision of the research.

The information provision of the research is based on information provided for the needs of the Ministry of Justice. In addition, the author has also used any publicly provided information by the Supreme Judicial Council, the Supreme Court of Cassation, Prosecutor's Office of Republic of Bulgaria, Ministry of Interior, Ministry of Culture, National Statistical Institute, EUROSTAT and etc.

Approbation of the research.

The research is approbated in front of 30 prosecutors and criminal judges from all over Bulgaria at the National Institute of Justice at the Supreme Judicial Council for the period 29-30.09.2014 (National Institute of Justice, 2014).

The research is also approbated through the public electronic and print media in the period 17-23 November 2014., together with the archaeologist prof. Nikolay Ovcharov and others.

# 1. Theoretical basis of the concept for opportunity costs of crimes against cultural-historical and archaeological heritage

In 1968., the Nobel laureate Gary Becker made some conclusions in his publication „Crime and punishment: an economic approach", which conclusions put the basis of the contemporary economic analysis of the criminal law (Becker, 1968):

- At the microeconomic level, criminals determine their actions through a rational approach on economic basis as they receive utility of their actions or inactions and the resources invested.
- At the macroeconomic level, through the regulation of costs of incentives for prevention or detection of crimes and the implementation of effective penalties, the damage, caused by the crime to the society could be limited to a pre-selected optimal level. .

According to Becker, *„the price of crime"* for the society includes: (1) public costs of the government bodies, who are engaged with counteraction of crimes and imprisonment of criminals, (2) private security costs, e.g.. for alarm systems, insurance and etc.., (3) the value of the damages, caused by the crimes, incl. missed benefits caused by the revenues loss or creating a product as a result of a murder, grievous bodily harm and etc. (4) missed benefits for family and relatives of the offender who are in his care as a result of his detention.;

The questions that rise, could be summarized as follows:
- Is it possible, the amount of damage to the society because of the crimes to be precisely determined through economic methods?
- What should be the balance between costs for incentives and costs for crimes?
- To which incentives and to which punishments, criminals react?
- What should be the amount of public expenditures for prevention and detection of crimes and re-education of criminals?
- Is it possible to be determined an optimal scale of the crime in relation to the marginal benefit for the society while maintaining the crime in certain limits, as well as the marginal costs, the elasticity of crimes in comparison to the measures taken for its counteraction?

*This study was published for the first time in 2015 only in Bulgarian. The current version is its first English translation.*

The author considers that the research of Gary Becker does not answer unambiguously to the abovementioned questions. For example, measurement of damage, caused to the society by the crimes is determined by the number of crimes, value of the property affected and direct missed benefits because of lack of product or incomes. However, it is not given a concrete answer to the question if moral damages could be perceived as economic ones and what their value to the individuals and to the society is. For example, how can be valued the suffering of a victim because of a rape and the long-term psychological problems she will have. And what about the loss of a child because of a car accident and how their parents will bear the loss.

On the other hand, the balance between incentive fees and costs for penalties depends firstly on the correct definition of the damages caused to the society and secondly on the institutional effectiveness of the judicial system. Does this mean in case of probation, the society will receive much more benefits due to society services as punishment and this method is more economically efficient rather than in case of imprisonment where the society's costs for living of the prisoner are higher. How could be calculated the reduced working activity of a victim with bodily injures, who has seen her robber to walk freely on the street as a probationary.

The decreasing marginal profit for the offenders as a result of the increasing possibility to be punished could influence the behavior of those offenders who are not risk-averse and are risk-neutral, but for those who are risk-averse according to a certain reference point of the incomes from crime activities could not be influenced from the behavioral economics viewpoint. The incentives, expressed in continuous increase of incomes from legal activity could also be inefficient, because it could be done through subsidizing the education of the potential criminal, but the incomes as a result of the higher education will be delayed in time until his graduation.

Is it possible the effectiveness of the judicial system to be measured as the number of crimes committed is divided to the number of sentences imposed? If it is assumed that there is a difference between crimes committed and crimes documentary registered in the judicial system by the police officers and the prosecutor's office and there have existed categories of crimes in which the victim is not an individual person but the entire society, then the effectiveness of the judicial system will be hardly identified.

Recent research in the field of measuring the economic dimensions of cultural-historical and archaeological crimes is that of Richard Freeman from Harvard University – "Economics of crimes" (Freeman, 1999).

Identifying that crime is a field of extreme behavior, which fact puts the economic analysis on test, Freeman considers the great amount of economic activities which are related to the security and internal order at the public and private sector or those sectors which are affected by the criminal activity. Based on the statements, the proposed methodological guidelines of Freeman are grouped in two directions: (a) determinants of crimes and (b) measuring the costs of crimes.

(A) determinants of crimes according to Freeman**:**

-*Opportunity for legal incomes;*

Low level of education, low personal skills, and experience to access the labor market, low legal incomes.

-*Risk propensity;*

Crime inherently is a risky activity; which fact is important in the decision making process. According to Freeman, criminal behavior is a subject of strategic games by police, criminals, and the society, examined through the viewpoint of the prisoner's dilemma. The risk is very often linked to the capture and imprisonment of criminals, which excludes the opportunity the criminal to commit new crimes while he is imprisoned and respectively to receive illegal incomes. At the same time, it prevents the criminal to have legal job and to receive legal incomes. In addition, it is more difficult for ex-prisoners to start a legal job and they are very often paid lower legal wages.

-*Non-monetary factors.*

For example, criminal street networks are sources of status and social control of individuals with criminal tendencies. It could be assumed that from the viewpoint of drug dealers, they prefer to be drug sellers rather than having a "socially humiliating" legal job where the level of wages are lower because of the lack of education or lower level of education.

Social interactions amongst potential criminals, potential victims as well as the judicial system are out of the scope of the price regulated system. The increased number of criminals could decrease the likelihood a single criminal to be caught and could incite others to commit a crime by giving an example. An increase in the number of police officers theoretically could lead to lower number of crimes. However, the time sequence of correlation between number of police officers and amount of crimes is a problem because it is very hard the number of police officers to be increased as fast as the increase in the amount of crimes and to be found a statistically significant feedback, synchronized in time.

*This study was published for the first time in 2015 only in Bulgarian. The current version is its first English translation.*

The author considers that the abovementioned determinants of crimes are mainly correlated to the identification of the supply curve of crimes and they influence the elasticity of supply. However, the determinants of demand (limit of social-tolerable level of crimes) are not included so as to be identified the potential point of equilibrium. These determinants will be examined in the following sub-point (B) „measuring the costs of crimes ".

(B) measuring the costs of crimes according to Freeman:
- Identification of the supply curve of crimes
- Participation in crimes

The criminal activity of a separate social group according to Freeman could be decomposed through various ways. One of the ways, which is perceived as a parallel analysis of the labor force supply, is the criminal activity to be decomposed to *supply of crimes per capita in non-institutionalized population* (CPP) in terms of *numbers of persons who commit a crime in the group* – criminal participation rate (CPR) and *number of active crimes* ($\lambda$):

$$CPP = CPR \times \lambda \qquad (1)$$

However, the reliability of data concerning active crimes ($\lambda$) is very often under doubt. They are rising some problems linked to the identification of groups of crimes such as multiple crimes or one continuous crime which influence the number of crimes reported. For example, the drug sale is a continuous crime or multiple crimes which depend on the number of drugs sold.

The equation (1) shows how the average number of crimes per person can be calculated. The equation is also appropriate for determining the various categories of persons: individuals who commit lots of crimes for the relevant period and individuals who commit incidental crimes (one to two) for the same period. In this way the level of criminal activity per a criminal is presented and it could be measured in terms of average number of crimes per a criminal. It could be measured also the number of crimes committed in relation to the expected decrease in the number of crimes because of detention.

Freeman makes the conclusion that the total supply of crimes varies according to the circumstances. From economic viewpoint, the decrease in incomes from crimes, related to the increased number of convictions could be compensated by increase in the incentives of criminals to commit new crimes – for example a drop in the amount of the legal incomes of the criminal, compared to an increase in his criminal incomes for the same period.

In accordance with the presented equation model (1) of Freeman, the author of the current study considers that:

First, Freeman's conclusion that data about number of active crimes ($\lambda$) is doubtful is correct and is also applicable for Bulgaria. Not all the committed crimes are reported to the Police. For example, very often, the victims consider that the committed crime is not so serious that the Police will take it into consideration and will help them (stealing clothes, bicycle, pets and etc.) or the victim is ashamed even though he/she is injured (in a fight, rapes or etc.). On the other hand, the Police does not register all signals for committed crimes. For example, crimes, where the victim is not an individual person, but the entire society. Such type of crimes are also crimes against the environment. In the cases of illegal cutting of timber, where great amount of trees are missing, sometimes the Police does not register the crime because it is clear that the criminal will be hardly revealed and this fact will influence the statistics concerning the number of crimes detected. The mechanism is almost the same in smuggling and tax evasion crimes. On the other hand, as there is not a specific victim, nobody will appeal this inaction to the prosecutor's office. This all means that the real number of committed crimes is much higher than the number of officially registered crimes by the Police, which number will influence the values of ($\lambda$).

Secondly, according to the author, the criminal participation rate (CPR) is related to the number of crimes and it does not take into consideration the different economic and social effect because of the separate criminal activities. For example, robbery causes more severe economic and social consequences compared to general theft, because it is related to violence and threats. This results not only in property problems but also in higher health costs and worse living conditions of the victim and his/her relatives because of fear and lack of peace. The same statement is applicable to many other categories of crimes, such as comparison between crimes against public and state authorities and domestic fraud. This means that the criminal participation rate (CPR) linked to the total number of crimes does not take into account the severity and public danger of separate crimes, thus giving unreal information about economic and social consequences of crimes and defining the focus of public counteraction.

The conclusion is that the correct ratio should be between number of crimes (in categories) and the severity of consequences (social and economic costs of the victim and the society). In this classification it is possible that one of the crime categories with low criminal activities could be much more dangerous than other category of mass criminal activities. In view of this, (CPP) as an indicator, considering the number of persons committing a crime, it presents the criminogenic situation or the criminogenic potential, but it does not lead directly to any economic conclusions.

- Economic context of utility for the offender

    In the prism of Freeman's standard economic decision-making model, people choose between criminal activities and legal activities, based on the utility of the relevant activity. If "Wc" is *profit of successful crime*, "p" is *likelihood the perpetrator to be detained*, "S" is *level of punishment* and "W" is *income from legal job*, people in the decision-making process will choose to commit a crime in a given period of time rather than to have legal job when:

$$(1 - p)U(Wc) - pU(S) > U(W) \qquad (2)$$

This equation highlights three types of effects:

First, it implies that in order the criminal activity to be chosen rather than the legal job, it needs to bring more incomes. If *p = O, U(Wc) > U(W)*, then Wc > W. The more price differences between "Wc" and "W", increases, the more the level of crimes stays constant. The successful crime must pay the sum, compensating the risk of capture.

Second, equation (2) means that the ratio to risk, measured by "U" curve, will influence the decision of committing a crime: those people who are afraid of taking risks will react to changes in the likelihood of criminals 'detention rather than in changes in the level of punishment, when the net revenues of crimes are fixed *((1 - p)Wc - pS - W)*.

Third, equation (2) shows, that main factors which influence the decisions of committing a crime – criminal versus legal incomes, likelihood the perpetrator to be captured, as well as the level of punishment – are inextricably correlated.

The author's opinion for equation (2) is that it consists of two activities in one-period model which treats crimes and legal jobs as substitutes. At the same time, the equation does not consider the number of committed crimes for a longer period and the opportunity costs of time, spent for planning and preparation of crimes. In the preparation of crimes there could be included also the costs of criminals for resources and what difficulties for the delivery of these resources they met.

- Defining the „market "supply curve of crimes

The individual decision of committing a crime is only the first phase of the economic analysis. To be supplied a separate crime, the equation of criminal activity should be aggregated amongst individuals and needs to present the supply curve of crimes:

| | | |
|---|---|---|
| CPP = *f(Wc .p. S. W)* | or | CPP =f(1 - *p)Wc - pS - W).p),*     (3) |
| CPR = *g(Wc. P. S. W)* | or | CPR = g(1 - *p)Wc - pS – W.p),*     (4) |

where the first ratio is the expected value of crime against legal job and „p" measures the risk of criminal activity.

According to the author, the main disadvantage of models (3) and (4) related to the supply of crimes is that they cannot explain some important components such as concentration of crimes in various geographic regions or changes in the concentration of crimes over time. Despite, models (3) and (4) have important influence over the efficiency of determining the % of prisoners in order the criminal activities to be reduced.

- Defining the supply line of crimes according to Freeman:
  From the viewpoint of supply, the curve represents a downward link between number of crimes and incomes of criminal activities. The victims of some crimes – for example drugs, prostitution, gambling are limited consumer goods, which consumers will buy less when the (function of "Wc") price increases.
  However the size of victims of crimes should be negatively related to "Wc" in terms of the expected revenue of the crime ((1 - *p)Wc - pS - W* in relation to the type of demand. One of the reasons is that the additional crimes may cause the society to increase „p" or „S", decreasing the profits of crimes (equations (2-4)).
  As the crimes financially and physically damage the victims, the government and individuals spend significant amounts of resources to prevent the crime. The optimization of the personal and social costs requires these activities to be done until reaching those point of the curve, where the marginal value of decreasing of crimes equals the marginal cost of relevant activity for preventing the crime.[iii]
  Theoretically, if all participants in the criminal justice system are optimized and there are no external effects of relevant prevention crimes actions then the scope of criminality to the society will be as "acceptable" as the society has considered. In the practice, however, it is arguable whether sanctions are more or less effective as a result of the social programs for prevention of crimes and the efficiency of the correlations amongst the efforts of the government and society for decreasing the amount of crimes.

In relation to the presented model by Freeman, the author considers that:

The increase of „p" or „S" from the society's viewpoint does not depend only on the number of additional crimes. This also depends on the socio-economic burden, distributed in categories of crimes and their structure. This will be determined not only by the direct financial consequences of crimes but also by the indirect economic effects, related to the reduced working activity of population due to fear of crimes (depopulation of neighborhoods and settlements) or failure to take economic incentives because for example the public procurements are rigged or business-competitors use corruption methods to reach their goals in disloyal way.

If we consider „p" and „S" separately, then „p" is *the possibility the criminal to be detained* and it is related to the police, judicial and private security resources, which the society spend for combating the crimes.

First, the analytical nature of statistically reported police expenditures (as far as they are publicly available) is very difficult. It is very hard to be identified what the police expenditures in various categories are because there is no data available about how many hours police officers have worked in separate categories. Moreover, the professional specialization of police officers does not correspond to the structure of the Criminal Code, concerning the types of crimes. In this way it is very difficult to be conducted an individual research about those categories of crimes with the greatest negative economic and social effect. It is not clear what type of information about these effects could be extracted from the police fixed costs, for example, costs for electricity, water supply and etc. and how these costs can be allocated amongst the various types of crimes.

The same statements could also be applicable to the judicial costs (prosecution, court, execution of sentence), which includes costs not only for magistrates and court administration, costs for legal aid and mandatory protection, but also costs for witnesses and jurors, which have opportunity costs for the time spent.

About the private security costs, there could be found some risks, concerning the statistical reliability of the preliminary data. From one side, it is very easy to be determined the costs for alarm systems, security systems, security, self-defense weapons. On the other hand, these costs can be related to negative external effects for the neighbors of robbed people and to direct thefts and attacks to neighbors, whose homes do not have alarm and security systems and home insurance. In this regard, the security costs depend not only on the willingness for prevention of the potential victim, but also on his/her financial resources.

Many other costs in this field are unclear. For example, changing the settlement of citizens in a better and safer neighborhood or city is related both to the escaping from the criminal environment or looking for a better job or better education for the children.

The public expenditures for prevention of crimes are hardly assessed effect. Especially those expenditures which have long lasting nature, such as for working with minors with hooliganism or for working with children at risk. On the other hand, effectiveness should be compared with the amount of expenditures so that even a separate program is efficient and the higher expenditures for its implementation are, the better they will be focused on strengthening the capacity of the prosecution or police offices.

Other costs, which should be considered are**:** opportunity costs of the victim as a result of police and judicial procedures, caused by the crime (lost working or free time of the victim, as well as the lost incomes and utility of the free time for caring for children, entertainments and etc.); the same statement is applicable for the employers of the victim, who cannot benefit from their employees at work when they are at the Court or in hospital; as well as a drop in the number of services provided at neighborhoods with high criminal activity and relocation of better shops and cafes in safer neighborhoods, which is directly linked to the provision of better paid jobs at these companies for the residents of these neighborhoods.

Cost affordability is also important but unclear component. The higher level of criminal activity could lead to an increase in the tax burden because of the need of over-funding for the judiciary to overcome the higher level of criminal activity, which could result in lower tax collecting and under-funding of the system.

When considering „S", which is *the level of punishment,* the author of the current study considers that „S" depends on capacity of places of detention, on the opportunities of the society to implement a good working probation system, as well as on the moral values of the society to efficiently impose public reproach and rebuke in case of not so publicly dangerous criminal acts. In this way, judges and legislators will take into consideration these opportunities of the society when impose and determine the detention for the criminal actions.

- Possibilities for targeted (preliminary determined) equation between supply of crimes and crimes constraint line

    The great advantage of imprisonment is that it removes criminals form the society so that they cannot commit new crimes. If we consider the great number of various crimes, committed by people with persistent criminal habits, then their imprisonment will have exclusively great

effect in reducing crimes. The reduced amount of crimes because of imprisonment, could be analyzed by using a demographic reporting framework. Arithmetically, if someone who commits 5 crimes per year in a separate neighborhood is imprisoned and no one replaces him, then the number of crimes in this neighborhood will drop with 5.

The market model says what exactly is missing and puts the focus on the additional information needed to be evaluated the benefits of imprisonment. According to the standard model of imprisonment it is implicitly accepted that the supply curve of crime is inelastic. When we have zero elastic supply crimes curve, then the inward displacement of the curve because of imprisonment will decrease the number of crimes. But if the supply of crimes has positive elasticity, then the effect of change will be less. In an ultimate case, infinitely elastic supply crimes curve means that imprisonment of one criminal "creates" another criminal or exceeds the speed with which the existed criminals commit crimes ($\lambda$), so that the imprisonment does not have an effect over the level of crimes. In terms of demand and supply, the impact of *increased imprisonment* ($\Delta l$) over supply of crimes is:

$$\Delta C = \eta \Delta l / (\varepsilon + \eta), \qquad (5),$$

where „$\eta$" is *elasticity of demand of crimes* and „$\varepsilon$" – *elasticity of supply of crimes*. In terms of equation (5), the effects of various strategies for criminal conviction could be evaluated. If we consider the effect of imprisonment in the context of the market model, we will understand that this effect is too modest to reduce the number of crimes.

In relation to the presented model (5) by Freeman, the author of the current study considers that:

This model (5) does not consider the number of unsuccessful attempts of criminals for committing a crime, which number may have increased no matter that the number of committed crimes is decreased. In this way it could be correctly defined the level of the real criminal activity.

In the model, it is also not considered the technology progress, which makes technically easier the committing of crimes and is related both to cheaper means for committing crimes and reducing the time for committing the crimes (for example, using a laptop and software for unlocking luxury cars and etc.).

- Measuring the benefits of reducing the crime.

> For the complete cost-benefit analysis of crime prevention needs to be calculated the marginal utility of each currency given for reducing the crime. However, it is hardly to be defined, because the value consists not only of the reduced monetary losses, but also of the reduced non-pecuniary losses for victims of crimes.
>
> The possibilities for determine the correct value of crime are based on the judicial jury's assessment of the non-monetary costs of the victim and the medical expertise of the damages of the victim, including also psychological ones. In all their problems, these estimates are arguably closer to the truth rather than the models, which are limited only to the stolen money.
>
> The skyrocketing populations of prisons and arrests, together with the increasing costs, raise many questions about whether "prison pays off". The answer to the question depends partly on the potential number of crimes, which the prisoner could commit if he were free and on the reaction of the others over the crime's margin to the imprisonment. It is very difficult to be measured the utility victims and society could receive because of the imprisonment of criminals. In each case, the highest cost of crime and imprisonment shows that if imprisonment is paid on a margin, even in less effectiveness of the alternative procedures in determining of the punishment – restraining order, e-surveillance, parole and etc., they will make sense.

In relation to the model presented by Freeman for measuring the benefits of reducing the crime, the author of the current study considers that:

He has not taken into consideration the geographic shift of crime; which fact compensate the effect of reducing the number of separate crimes in one town or region on national level.

Ha has not also taken into consideration the type of displacement of crime, which fact is in response to the concentration of police' efforts to prevent relevant category of crimes, however the limited resources of the police do not prevent criminals from committing other types of crimes. For example, street robberies can be decreased if more police officers are patrolling amongst the streets, but if this fact leads to an increase in the number of documentary or telephone frauds, then the effect will be compensated.

Another unaccounted effect is the multifaceted action of the separate measures for reducing the crime. One incentive of the society (for example, the presence of more police officers at streets) could lead both to reducing the crimes against person (murders, bodily harm), and to reducing the crimes against property.

*This study was published for the first time in 2015 only in Bulgarian. The current version is its first English translation.*

The publication of Richard Freeman, „Economics of crime ", despite its disadvantages, proves that with the methods of economic analysis the process of public crime counteraction could be managed by comparing costs for prevention of crimes and costs for detention of crimes with the savings of the police, justice and prison authorities, as well as with the savings of the real or potential victims of reduced criminal activity or increased detection of crimes.

In this regard, the society and the state authorities could choose from the most efficient policies by comparing them one to another as they allocate the constraint budget resources in the best way to counteract the crime. The efforts will be concentrated not only to reducing the total number of crimes, but also dealing with those categories of crimes which result in the most negative economic and social effects for the society, even though their nominal number is not so high (domestic crimes for example).

As a result, the answer to the question – what is the correct level of public resources compared to the level of crime which the society could endure? – is of key importance. In addition, another question will be raised – what needs to be the ratio between: *prevention-detention-mitigation of negative consequences* of crimes, compared to the chosen level of costs?

In the model of Freeman, it not sufficiently incorporated the economic concept for „opportunity costs ", so as the costs for crimes to be measured more correctly. This is possible to be done, for example through determining the alternative usage of the police and judicial authorities for reducing the crime or the alternative usage of the material and financial resources, which are saved from criminal encroachment or because of reducing the crime, these resources have not been spent for mitigation of its negative consequences.

In the model of Freeman, the costs of crime are not sufficiently refracted through the concept of *„transfer costs "*. Insofar a theft is an unwanted transfer of property from victim to the criminal, then it is often related to some type of concealment of the stolen item and its subsequent legalization (through false documents or other type of "laundering"), which is associated with its own costs. Insurance is also type of transfer costs (rather anti-transfer) for the potential victims.

In this regard, the "transfer costs" of crime are not equal to zero, as it is stated in the Freeman's model and it is needed, they to be considered.

The possible application of Freeman's model for determining the opportunity costs of crimes against cultural-historical and archaeological heritage will be discussed in the next chapter of the current study.

*This study was published for the first time in 2015 only in Bulgarian. The current version is its first English translation.*

## 2. Model for determining the opportunity costs of crimes against cultural-historical and archaeological heritage based on demand

In accordance with the conclusions made in the previous chapter and based on the model of Freeman, it could be determined the opportunity costs of crimes against cultural-historical and archaeological heritage but on a demand-based modification.

The revised model should consider the determination of the total economic value of objects of cultural-historical and archaeological heritage. Having in mind this value, it is also needed to be considered the expenditures society spend, which are reciprocal to the negative effects of these types of crimes, determining their demand constraint line.

The revised model should also consider the specific factors that influence the utility of the criminal, motivating his criminal behavior and shaping the supply curve.

The result of the implementation of the revised model should present the possibilities for reaching a targeted equation between supply of these types of crimes and their demand constraint line.

Measurement of the likelihood the criminal to be apprehended *(p)* in crimes against cultural-historical and archaeological heritage in terms of demand

If we are following the Freeman's model, then the likelihood the criminal to be apprehended will depend on the expenditures the society spend for counteraction of these types of crimes.

These expenditures will depend on the:

-size of the damage, caused to the society;

-number of crimes against cultural-historical and archaeological heritage;

-budget constraint of government, companies and individuals.

The analysis needs to start with the identification of the most unclearly studied in the literature factor – the size of the damage caused to the society by these types of crimes:

For measuring the damage of these types of crimes, it is needed to be determined the total economic value of object of these crimes (movable cultural-historical and archaeological values).

On the first place, this is the value of direct usage.

When determining this value, the value, created by the direct usage of the object and the additional monetary value, created by the secondary economic activity, caused by the direct usage of the object, should be considered.

In determining the value created by the direct usage of object (direct monetary value), there could be applied various types of approaches:

The normative approach should be based on the Ordinance on the procedure for identification and maintenance of the Register of movable cultural values (Ordinance N-3/2009). The positive thing about this approach is that it is based on a binding legal act, implemented by the government. A disadvantage of this ordinance is that according to the procedure it determines the nature of the object – if it is a cultural value or not, as well as if this object is a cultural value of high level (national heritage), however the direct monetary value of the object is not identified. For this monetary value, there could be made some assumptions in terms of the level of the relevant cultural value, but they are in a wide range. Even the monetary values of cultural and archaeological objects, which are national heritage could vary in difference of ten to thousand times.

Another opportunity for application of the normative approach is the Ordinance on determining the amount of remuneration of persons who have handed over objects in accordance with art. 93 of the Law on Cultural Heritage (Ordinance N-2/2012). The positive thing in this legal act is that it is related to the determination of the exact monetary value. The negative thing is that this monetary value has a fixed ceiling of 2500 Euro (art. 3, p. 2), which ceiling is extremely insufficient because the real price of cultural objects from Bulgaria can reach values such as tens of millions of euro per item.

Consequently, having in mind the disadvantages of the normative approach, it could not be successfully applied for determining the real value of cultural and archaeological values in terms of judicial and pre-trial procedures, as well as for conducting scientific research to determine the size of the public damage, caused by such types of crimes.

The market approach could be based on the auction, insurance, or other known market prices of objects from the same era and with similar identification parameters, if the studied object has had a previous analog. This approach is sometimes used in judicial and pre-trial procedures in the Bulgarian case law.

For example, the value determined as auction value (initial or sell value) at international auctions, if the objects have analogs (similar objects have already been sold) and if they are evaluated by appraisers according to the methodology of the international auctions. Alternatively, the insurance value of objects could be determined by the export price of similar objects (if they have analogs) to international exhibitions or if they are evaluated by appraisers according to the methodology of known (famous) insurance companies. Also, if the aforementioned opportunities for determining the value of cultural and archaeological objects are missing, then it is possible to be used the price at the so called "black market" if the object has an analog and this price is well-known (ads in personal websites and others). As last opportunity, there could be used the value, based on the revenues of tickets selling at museums for a relevant period.

On second place, this is the additional monetary value.

It is secondary value because of the direct usage of the object. This value could be measured by the:
- revenues from restaurants and souvenir shops at museums at which these objects are exhibited,
- revenues from hotels and transport for visitors from other town or city or country who have come to see the exhibited object at the museum,
- revenues from advertising usage of the object, for example as labels of wine bottles or tv commercials.

The measurement of the parameters mentioned above, is done as a difference between the time before the object was exhibited and the time after the exhibition of the object is ended at the relevant neighborhood or town/city.

On third place, this is the indirect value or the value of non-use of the cultural-historical and archaeological value.

The definition of value of non-use was firstly introduced by Barton Weisbrod in 1964 in relation to the cost-benefit analysis (Weisbrod, 1964). For the period of the 60 years of XX century to nowadays this value is mainly related to studying the values of components of the environment. However, there are many studies which examines the application of this value in the context of the economics of culture. The publications of Throsby and Brooks will be used as theoretical basis for examining the indirect value of movable cultural-historical and archaeological values in the current research (Throsby 1984; Throsby 2010; Brooks 2002). There are many publications in the context of economics of culture, however they are empirically oriented and could be analyzed in future studies of the author (Lundhede, Bille and Hasler, 2013).

By the indirect value, there could be measured the social value of the objects of cultural-historical and archaeological heritage, considered as an effect over the public welfare. This is the value of benefits for all members of the society, beyond those who have had direct contact with the object.

According to Brooks, the indirect value of the cultural goods includes:

- *Value of existence.*

    Even though, some people do not consume directly the relevant cultural good, they could evaluate its own existence.

- *Optional value.*

    People who do not consume a relevant cultural good could give positive value of the option to become future consumers of this good and in this way to favor its protection.

- *Educational value.*

    Some cultural good could influence the consumers and other members of the society in intellectual and economic manner.

- *Prestige value.*

    Some cultural goods could create prestige of the region of their origin or exhibition.

- *Donation value.*

    Local and foreign consumers can benefit from the expected consumption of the cultural good by future generations.

According to Brooks, the content of the indirect value of cultural goods assumes that not all benefits of the art could be consumed by the today's society. In case of historic prevention ("cultural heritage"), the object will embrace both the unborn generations and generations who are living today. Having in mind that in most cases unborn generations are often very important for the arguments of the public policy for art funding in the context of sustainable development, then Brooks focuses his research on the readiness of the public to fund the cultural values prevention.

Similar thesis about the content of indirect value of cultural goods could be found in Throsby's publications. According to him this value is decomposed to aesthetic, spiritual, social, historical, symbolic, and authentic value. In addition to the components, Throsby examines "*value of national identity*", determined as "the feeling of being Britain". He puts the basis of this value on the subjective-socially shared experiences in a community, which fact makes very difficult the determination of curve and its individual and aggregated utility in monetary terms. Moreover, this value summarizes the elements of the other listed by him values (spiritual, historical, symbolic, authentic), which makes difficult the measurement of its share in the frames of the indirect value of cultural goods. Despite this, the value is very useful for determining the public damage, caused by crimes against cultural-historical and archaeological heritage.

The author of the current research considers that the elements, presented by Throsby and Brooks as content of indirect value of cultural goods do not cover all possibilities. Their studies examine more contemporary art (predominantly fine art, music, fiction, etc.), which covers periods about the human history for which there is sufficient historical information.

In the cases of Bulgaria, Italy, Greece, Spain and other countries, which are rich of antique (archaeological) cultural heritage where many cultural goods have the status of world cultural heritage according to the UNESCO classification and that is why they are global public goods, these elements are not sufficient for the determination of the indirect value of cultural goods.

The specific thing in those type of cultural and archaeological values is that as public goods and in their public consumption, their real market value is not considered. For example, the access to those goods could be free or state subsidized which means that their market value may not be in the form of enter thickets or coupons. They may not have analog which is traded on the markets so far. In this aspect, the determination of parameters of demand curve, as well as the relationship to the consumer surplus is very difficult.

If we accept Brook's thesis that all benefits of cultural values could not be utilized by the current society and this leads to the necessity of their prevention, then they will be some form of external costs, imposed by the future generations to the current generation. Consequently, the funding of conservation and prevention of cultural values raises the question about the degree of altruism of the current generation and the lack of which determines the percentage of potential crimes against cultural-historical and archaeological values. This potential could be determined by the willingness to pay model and will be discussed below.

Initially, there will be made an analysis of the elements of the indirect value linked to the cultural-historical and archaeological values, presented by Throsby and Brooks so that to determine the possibility these elements to be included in a mathematical model, measuring the public loss as a result of the crimes, based on the total economic value of the object.

The first element is *„value of existence" of* the cultural-historical and archaeological value. The author of the current research shares the opinion of Brooks and Throsby (presented above) defining this value as benefit of the very knowledge of the existence of a separate cultural good without the need, this good to be visited now or in the future. The key moment in the definition is that the individual receives benefits, no matter that he/she will visit or will have direct contact with the cultural good. This satisfaction is also received by the knowledge that the cultural good exists and it will continue to exist and it is related to the individual efforts this good to be preserved in the form of additional public tax or private donations. Consequently, it is possible this value to be measured through the contingent valuation method.[iv]

The big problem in measuring this type of value is that the benefits it provides could be consumed by the so called "free riders" who assume that there are enough individuals and organizations who will pay for the continued existence of the public good and there is no need for them to allocate financial resources for the prevention of the cultural good.

If this issue is examined in terms of crimes against cultural-historical and archaeological heritage and the statements above are considered, then we could conclude that there is no universal measurement of the values of the existing good. Local people in regions, rich of archaeological findings and at the same time with high unemployment rate (for example Vidin region, the zone of archaeological reserve „Nikopolis ad Istrum") are mainly interested in the amount of their fixed costs (for food, housing, medical cares). As a result, most of these people could be „treasure hunters" which fact leads to destruction of the good at the region or obstruction of public knowledge and access to it.

This value implies considering a long-term perspective and because of the low quality of life, accompanied by low economic development, the establishment of the cultural good in the region of its origin or its exhibition will be very difficult. This also influences the determination of the supply curve of crimes, as the potential criminals of these crimes in the short-run would find it difficult to be convinced not to commit the crime with arguments for value of existence of the cultural good compared to the arguments for hunger and poverty.

On the other hand, the exclusion of this value from the analysis is incorrect. Many people have not seen the so called „Varna Gold Treasure", which is one of the oldest human processed gold in the entire world.[v] At the same time, however, most of them are proud to the fact that this treasure was discovered in Bulgaria and shows that in fifth century B.C. in these lands, there were state formations and developed civilization of which they are heirs. This knowledge brings people some degree of utility without ever having to see the treasure even on picture.

Therefore, this value will be considered as an element of the indirect value of cultural-historical and archaeological goods in the current analysis.

The second element of the indirect value is the „*optional value*" of cultural-historical and archaeological goods. According to the definition given by Brooks and Throsby, presented above, the „optional value" is the evaluated positive value of individuals to become future consumers of the cultural good and to benefit from it, even though they are not direct consumers at the moment.

The origin of this term is related to economics of public welfare and corresponds to the willingness to pay for prevention of an ecologic or transport asset while retaining the minimal opportunity this asset to be ever directly used, which is the difference with the value of existence of the cultural good, where its direct use is not considered (Brookshire, Eubanks and Randall, 1983).

Despite this, both values are close one to another and there are parallels between them, which make risky their usage at the same equation for determining the total indirect value of the cultural good. The main similarity is related to the willingness in both cases to be paid some money to preserve the cultural good as the reasons for that could be complex.

It is very difficult to distinguish the utility received by the individual at the moment of understanding that such type of cultural good is existing and the utility of knowledge that the individual has the opportunity if he decides to visit this cultural good in the future. However, it is correct these components to exist as separate ones, as various individuals may decide to give zero grade to the one component and positive grade to the other.

Therefore, this value will be perceived as an element of the indirect value of cultural-historical and archaeological goods at the current analysis.

The third element of indirect value is *„Educational value"* of cultural-historical and archaeological goods. According to the definition given by Brooks and Throsby, the "educational value" of cultural-historical and archaeological goods is their ability to create intellectual and economic impacts amongst consumers and society.

The educational value of these goods consists of values transmission (social, moral, esthetical), which are the basis of individuals behavior and creates their character in the long run, as well as the public welfare. These values help individuals to be socially responsible, to strive for development, to look for decisions for the raising problems through critical thinking and imagination, to talk with the others and think ethically, to show creativity and interest in knowledge. This helps individuals to develop innovative and leadership qualities and to become better members of the society, contributing to the public welfare. By developing rich cultural individuality, these members of the society will better express their qualities finding meaning in their own existence and in the development of the human civilization.

As a subtype of the educational value it could be examined also the „scientific value". It is related to the extraction of scientific information from the national, European or the civilizational identity linked to relevant historical periods. The science has an important role in understanding nature and universe. For example the Thracian sanctuary at village of Tatul (called „Temple of Orpheus"), which is a megalithic complex dating back the period of XV-XIV century B.C., is very precisely oriented to the celestial bodies and cosmic movement mainly of the Sun, which presented functions in the Thracian rituals (Fol, 1998). The sanctuary and objects, found in it, give scientific information about the level of Thracians achievements and accordingly about the separate stages of the human civilization development.

Consequently, the educational and the scientific value as a part of it, will be considered as elements of the indirect value of cultural-historical and archaeological goods in the current study.

The fourth element of the indirect value is the „*Prestige value*" *of* cultural-historical and archaeological goods. According to the definition, given by Brooks and Throsby, the "prestige value" of cultural goods is their ability to create prestige for the region of their origin and exhibition.

Practically, in regions where there have been found or exposed popular cultural-historical and archaeological values, the citizens will receive benefits from this fact even though they have not visited these objects. The citizens will receive benefits under the form of reputation and respect by individuals who live in other towns or cities and who know about the prestige of these values. This is related to the social status of individuals who receive benefits from the prestige value of the good no matter what the personal qualities of these individuals are. This fact gives individuals higher social evaluation in their initial contact with other people, because of the prestigious quality and reputation of the cultural-historical and archaeological good which is found or exposed in their living region.

Consequently, the prestige value will be considered as an element of the indirect value of cultural-historical and archaeological goods.

The fifth element of the indirect value is the „*Donation value*" of cultural-historical and archaeological goods. According to the definition, given by Brooks and Throsby, the "donation value" of cultural goods is the likelihood, consumers who are familiar with the existence of the good to extract benefits from the expected consumption of the cultural good by the future generations.

First, unlike the optional value, the donation value is related to the potential visit or observation of the cultural good, which even if it does not happen, it has always been an alternative to the optional value. On the contrary, the donation value is related to the performance of some action for conserving the cultural good (funding, hiring security officers) in order it to reach to the

*This study was published for the first time in 2015 only in Bulgarian. The current version is its first English translation.*

future generations so that they will enjoy it, study it or receive any other benefit from it. One may think that at the moment this cultural good has no value or utility, but for the future generations it will have (for example, it is known that many paintings of a famous painter in the past increase their value over time and future generations appreciate them more).

Consequently, the donation value will be considered as an element of the indirect value of cultural-historical and archaeological goods in the current study.

Throsby defines as elements of the indirect value also the following ones:
- Aesthetical value,

  This value however is an element of the direct value, because the benefit received as a result of the direct contact of the individual with the good could be measured through the market mechanism as a price element of the museums tickets.;
- Spiritual value,

  This value is however an element of the educational value and its consideration as a separate element could result in double reporting, except in cases the cultural good has also religious nature.
- Social value,

  This value, however, is also element of the direct value, associated with secondary economic activity reflected through restaurants, hotels, and other places for social contacts, organized in and close to the museum where the cultural good is exhibited.
- Symbolic value,

  This value is also an element of the existence value of the cultural good, insofar it has no direct relation to the symbols of statehood or direct historic events linked to the establishment and affirmation of the state and nation.

Consequently, these values will not be considered as elements of the indirect value of cultural-historical and archaeological goods in the current analysis.

The measurement of values accepted in the current analysis and which present the benefits of the indirect consumption of cultural-historical and archaeological goods, could be done through many methods, one of which is the *Contingent valuation*. Other possible methods are „selective modelling", „hedonistic pricing", „methods of the transportation costs" and multi-criteria analysis. These methods could complement the measurements, done through the method of contingent valuation and could be combined with the method of multi-criteria analysis, as well as with non-economic quality methods related to museum, archaeological and cultural backgrounds.

The advantage of contingent valuation method is that it is applicable mainly in terms of environment and transportation (Harris, 1984). There are some publications where this method is used for analysis in terms of culture despite its disadvantages - it is expensive, time-consuming, it is related to distortion of the questionnaires and incentives to answer, lack of information by the respondents, wrong summarization and etc., (Lambert, Saunders and Williams, 1992).

According to Freeman's model, the likelihood criminal to be detained is based on the expenditures of the society to counteract of this type of crimes, which depend on three components, examined in the first section of the current research – size of the damage for the society which these crimes cause through determining the total economic value: direct + indirect, which has been analyzed so far.

Considering the other two components, the following is important:

- Number of crimes against cultural-historical and archaeological heritage.

  Number of crimes against cultural-historical and archaeological heritage determines the willingness of the state to counteract through establishing specialized bodies (for example specialized departments at General Directorate „Criminal Policy" at Ministry of Interior and State Agency National Security; General Directorate "Inspectorate for Cultural Heritage Prevention" at Ministry of Culture) or level of specialized workload of the general state bodies for counteraction of crimes (prosecution, security police and etc.). In this case the reliability and homogeneity of the reported statistic data by the police is of great importance. Some distortions are possible, regarding the reporting of higher detectability through incorrect or incomplete registering of new cases of crimes in periods, coinciding with different government mandates. It would be correct, such type of activity to be covered by judicial statistics. Unfortunately, at the moment, the judicial authorities use the statistical information for new cases, registered by the police and do not control the gathering of information through conducting sociological surveys at risk areas and groups.

- Budget constraint of government, companies, and individuals.

  The budget constraint of government, companies, and individuals in terms of counteraction of these types of crimes is determined mainly by the understanding of the importance and significance of cultural-historical and archaeological goods. The significance of these goods so far, was determined by the economic value, related to their direct consumption and measured through the creation of primary and secondary economic activity without taking into account the value based on the indirect consumption of these goods. Only their symbolic value is considered without determining its number, mostly when these goods are related to the symbols of state or nation. In similar cases, the government provides more serious security of the objects (for example National historical museum etc.).

The general macroeconomic rules for fiscal and budget policy for the government and the microeconomic rules for budget constraint of companies, non-profit organizations and individuals, owning collections apply here.

The conclusions made in the previous section of the study in relation to the Freeman's model applied to crimes against cultural-historical and archaeological goods and examined from the viewpoint of the demand of crimes, could be summarized as follows:

According to Freeman, individuals choose between criminal activity and legal activity based on the benefits they receive, and they will choose to commit a crime in a relevant period rather than find a legal job when:

$$(1 - p)U(Wc) - pU(S) > U(W) \qquad (2)$$

As it was mentioned in the previous section of the current research, the criminal will choose to commit a crime rather than do a legal job, only when the income he will receive is higher. If *p = O, U(Wc) > U(W)*, then Wc > W. This will happen only if there is no risk of detention of the criminal (*p = O*), which is specified in Freeman's method.

The author of the current study considers that if the method of „total economic value" of crimes against cultural-historical and archaeological heritage is used, then:

*p* = BC-TEV

where:

BC is the budget constraint, as for the government it depends on the macroeconomic indicators such as the fiscal ones and the sensitivity of the society to the amount of crimes, related to crimes against cultural-historical and archaeological heritage.

TEV is the „total economic value" of damage, caused to the society through the illegal impact over the cultural-historical and archaeological good – object of the crime. Where:

TEV= direct value + indirect value

- Direct value = value created by the direct consumption of the good and the additional monetary value, created by the secondary economic activity, caused by the direct consumption of the good and measured by the normative or market approaches, examined in the beginning of the current section.

- Indirect value = *existence value + optional value + educational value + prestige value + donation value*

The reporting of the "total economic value" of damage, caused to the society by crimes against cultural-historical and archaeological heritage, will force the society to increase the amount of expenditures for counteraction of these crimes.

The higher the risk of detention, the higher will be the price of illegal selling of cultural goods at the black market. As a result, the demand of these goods will decrease and the illegal incomes of criminals will also decrease, as well as their number. Then the government due to its fixed budget, because of the budget constraint of police officers, prosecutor officers and other employees will be able more efficiently to counteract to the reduced number of criminals in this field.

Of course, these criminals will not disappear, but they will orient their efforts to other fields (e.g. domestic thefts), but not all of them. This will further reduce the number of criminals in these types of crimes.

## 3. Empirical analysis of crimes against cultural-historical and archaeological heritage in Bulgaria

The purpose of the empirical analysis of crimes against cultural-historical and archaeological heritage is to be determined the parameters of *risk of detention* ($p$) for the needs of conducting a study for contingent valuation for Bulgaria and to test the model, presented in the previous section.

As it was mentioned in the introduction of the current publication (paragraph "information provision of the research"), in the conduction of the research, there will be used real statistical data provided by the state authorities especially for the needs of the research, as well as publicly available reports and analysis of the state authorities in the Internet. The available data is reliable and homogeneous as for the studied period, the legislation in this field is amended only once (in 2009) and consequently the information will be efficient enough for the needs of the research. As the study deepens and continues, more reliable and detailed data will be provided.

*This study was published for the first time in 2015 only in Bulgarian. The current version is its first English translation.*

When determining the parameters of *risk of detention* (*p*) it is necessary to be examined the relationship between data about the potentially committed crimes against cultural assets and the number of initially pre-trial procedures of these types of crimes.

This will give an idea about the level of counteraction of these crimes by the state control authorities, mainly Ministry of Interior and Ministry of Culture, including level of information about the committed crimes, difference between alleged crimes and officially registered ones by Ministry of Interior crimes.

Fig.1: Level of counteraction of crimes against cultural-historical and archaeological assets

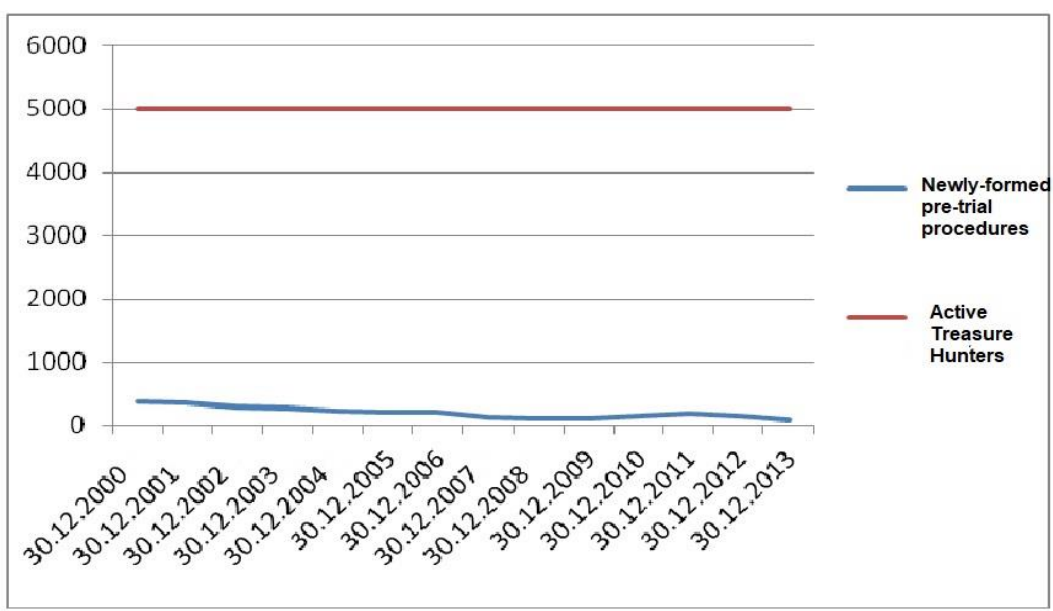

Source: author's own analysis, based on expert evaluations of representatives of pre-trial proceeding authorities and archaeologists

According to a published research, conducted by the Center for the study of democracy (2007) the professional treasure hunters are „several thousand people", but the incidental treasure hunters are 100 thousand people (CSD, 2007).

According to the expert evaluations of representatives of the pre-trial procedures' authorities and archaeologists, made for the needs of the current study, the active treasure hunters in Bulgaria are approximately 5 to 7 thousand people.

At the same time, the registered crimes by the Ministry of Interior for the period of 2000-2006 vary between 368 and 206 and for the period 2007-2013 they vary between 130 and 86.

Consequently, at Ministry of Interior they are registered under 1% of the committed crimes against cultural-historical values. This is also confirmed by the fact that great amounts of cultural-historical values are sold on the Internet (through the platform eBay) or have been illegally exported from Bulgaria and after that reimported from foreign countries thanks to the international legislation and police cooperation amongst countries. For example, in 2011 only one country (Canada) returned to Bulgaria 21 000 illegally exported antique coins (Bulgarian National Television, 2011).

Every year, most of the seized objects – cultural values, which have not been seized on the territory of the country, have been returned to Bulgaria by the state authorities of other countries.

Main reasons for not registering 99% of the committed crimes against cultural-historical values by Ministry of Interior due to:

- Lack of a real victim of this type of crime, as the victim here is the entire society compared to the domestic thefts where the main victim is a specific person and his family. This leads to lack of signals to the Police stations and low level of public control over the reported statistical data by the Police.

- The willingness of society to have higher level of detectability of crimes against cultural-historical values. As those types of crimes are more difficult to investigate and they are more difficult to prove, only cases which have convincing evidence on the first stage are registered.

There could be listed many other reasons, but the difference of 99% between the actually committed crimes and registered crimes by Ministry of Interior, means that the police intelligence does not work properly and the chance these crimes to be detained is under 1%, which is also indicated by *(p)=1%*

On the next place, it is necessary to be determined the level of detection of these types of crimes, measured by the number of officially registered cases by the Ministry of Interior. As determinants of the detection of crimes will be used 1. Number of pre-trial procedures submitted to the Court, 2. Number of convicted criminals and 3. Number of efficiently sentenced and prisoned criminals. The last determinant should carry the greatest weight because it is related to the risk of perpetrator's inability to commit new crimes, to deprive him of the opportunity to find a legal job while he is imprisoned and after serving the prison sentence. These effects will determine the *risk of detention (p)* of criminals and taking decisions by the potential criminal whether to commit such a crime in the future.

*This study was published for the first time in 2015 only in Bulgarian. The current version is its first English translation.*

Fig.2: Level of detention- in total for the crimes against cultural-historical and archaeological values

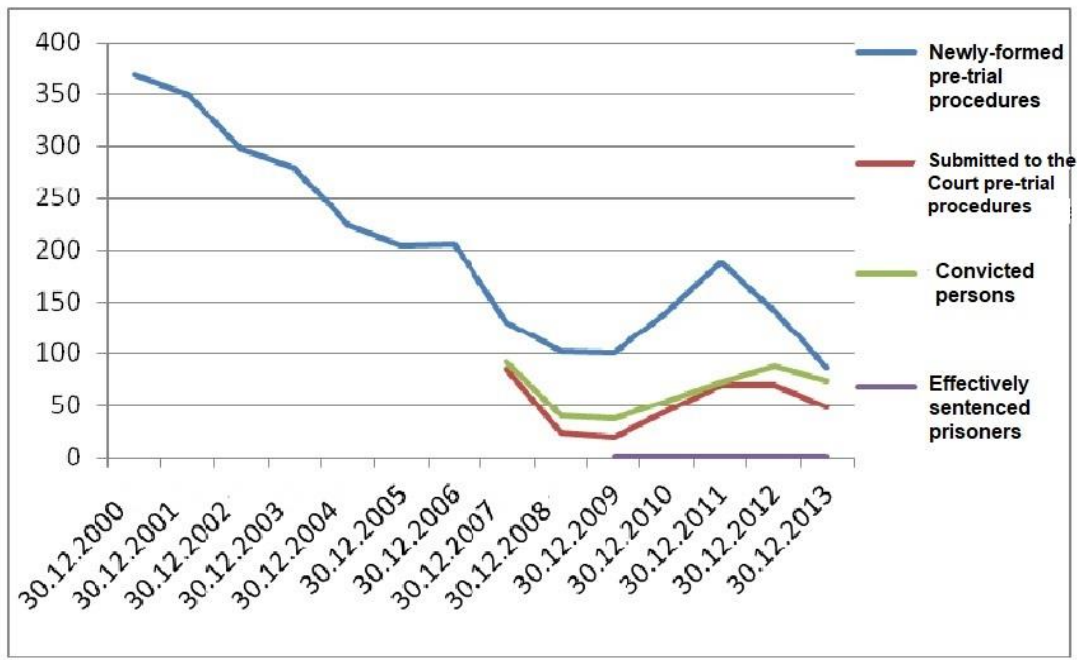

Source: author's own analysis, based on statistical data provided by: Ministry of Justice, Ministry of Interior, Prosecutor's Office of Republic of Bulgaria, Ministry of Culture

According to the data, showed on figure above, 60% of the registered crimes by Ministry of Interior are submitted to the Court. As in the separate cases more than one person is detained – the imprisoned people for this type of crimes are approximately 60 persons per year. However, only 1 person per year is with an effective sentence of imprisonment for all the committed crimes for the entire country.

This once again proves that the determinants need to be allocated in accordance with their weight, as the effectively sentenced imprisonment to the perpetrator should have the greatest weight amongst the others, which according to the data on the figure above indicates *(p)=1%*

The next step of the study is to be analyzed separately the types of crimes, related to cultural values. In this regard, there will be indicated whether the counteraction of the state and society to some types of crimes is stronger and as a result how to be allocated the public resources to efficiently counteract to those crimes.

*This study was published for the first time in 2015 only in Bulgarian. The current version is its first English translation.*

The first type of crimes which will be examined is the crimes related to the search and discovery of treasure containing cultural value (art. 208 of the Criminal Code). These crimes are only 10 % of the total number of officially registered crimes of this type.

Fig.3: Level of detention – crimes related to search and discovery of treasure containing cultural value (art.208 of the Criminal Code)

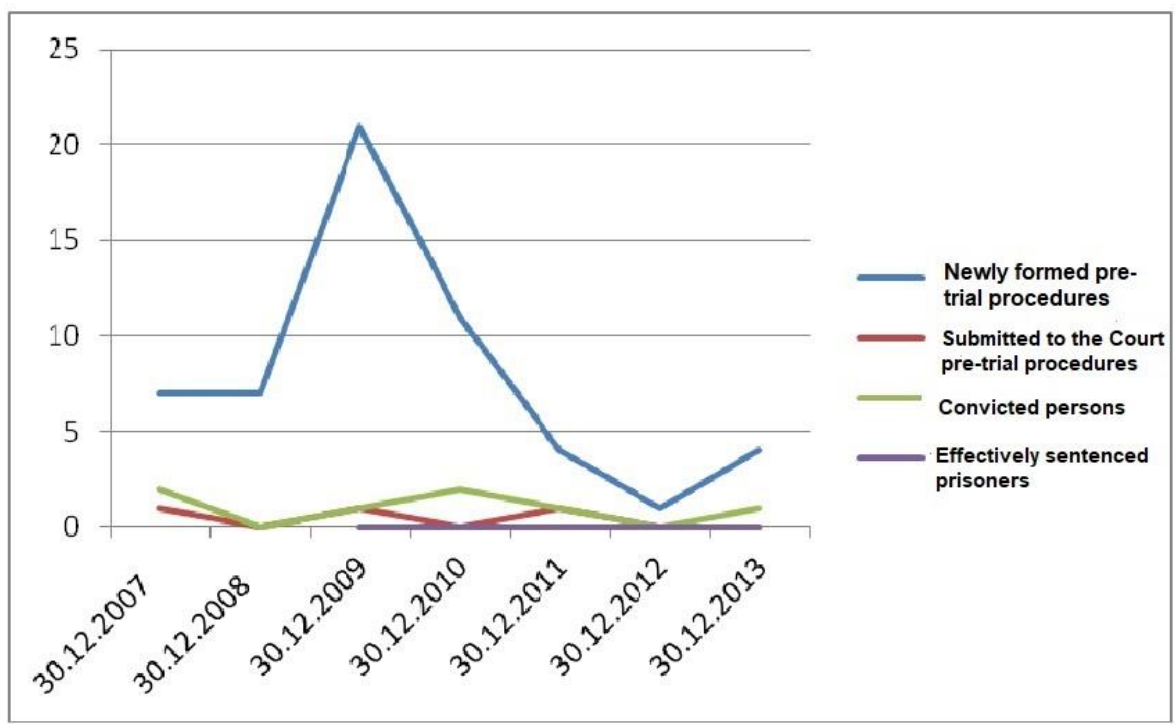

Source: author's own analysis of statistical data, provided by: Ministry of Justice, Ministry of Interior, Prosecutor's Office of Republic of Bulgaria, Ministry of Culture

What is impressive on figure 3 is the sharp difference between the number of registered crimes by Ministry of Interior – 10 per year, the number of submitted pre-trial procedures to the Court – 1 per year, the number of convicted persons – 1 per year and the lack of effectively sentenced prisoners.

This indicates (*p*)=0.1%

The second type of crimes are those related to the excavation and active investigation of archaeological objects (classical treasure hunting). These crimes are 30% of the total number of the officially registered crimes.

*This study was published for the first time in 2015 only in Bulgarian. The current version is its first English translation.*

Fig.4: Level of detention – crimes related to classical treasure hunting (art. 277a of the Criminal Code)

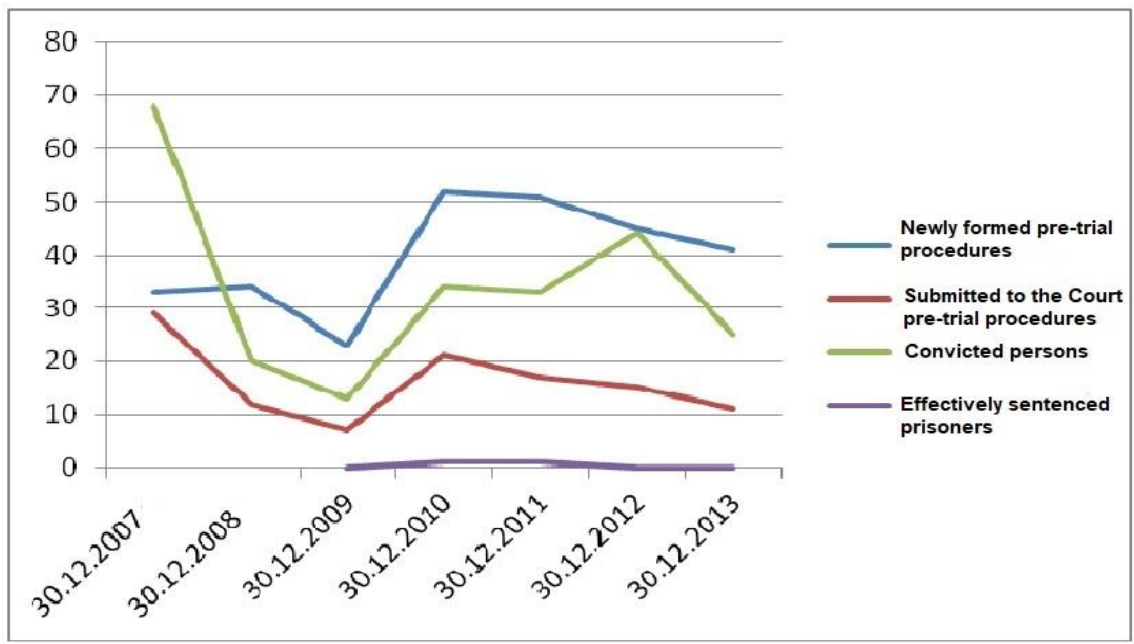

Source: author's own analysis of statistical data, provided by: Ministry of Justice, Ministry of Interior, Prosecutor's Office of Republic of Bulgaria, Ministry of Culture

Seen from the figure above, half of the registered crimes of this type by the Ministry of Interior are submitted to the Court for examination. At the same time, the number of the convicted persons is the highest in comparison to all convicted persons of other crimes of this type. This shows that the counteraction of the state and the government is concentrated in this field. A weak point of this strategy of counteraction is that the persons who commit such crimes stay at the lowest level of this criminal hierarchy in the treasury hunting process and they are called "diggers". As long as the middle level of "dealers" and top level of "patron-exporters" at the criminal hierarchy are not affected, then ordinary diggers could be easily replaced, especially in regions with high unemployment rate and existence of many archaeological objects, mainly in the region of South-West Bulgaria.

Only 1 person per year is effectively imprisoned because of all these committed crimes.

This indicates *(p)=1%*.

The third type of crimes are those related to concealment of cultural and archaeological values. These crimes are 30% of the total number of officially registered crimes of this type.

Fig.5: Level of detention – crimes related to concealment of cultural values (art. 278 of the Criminal Code)

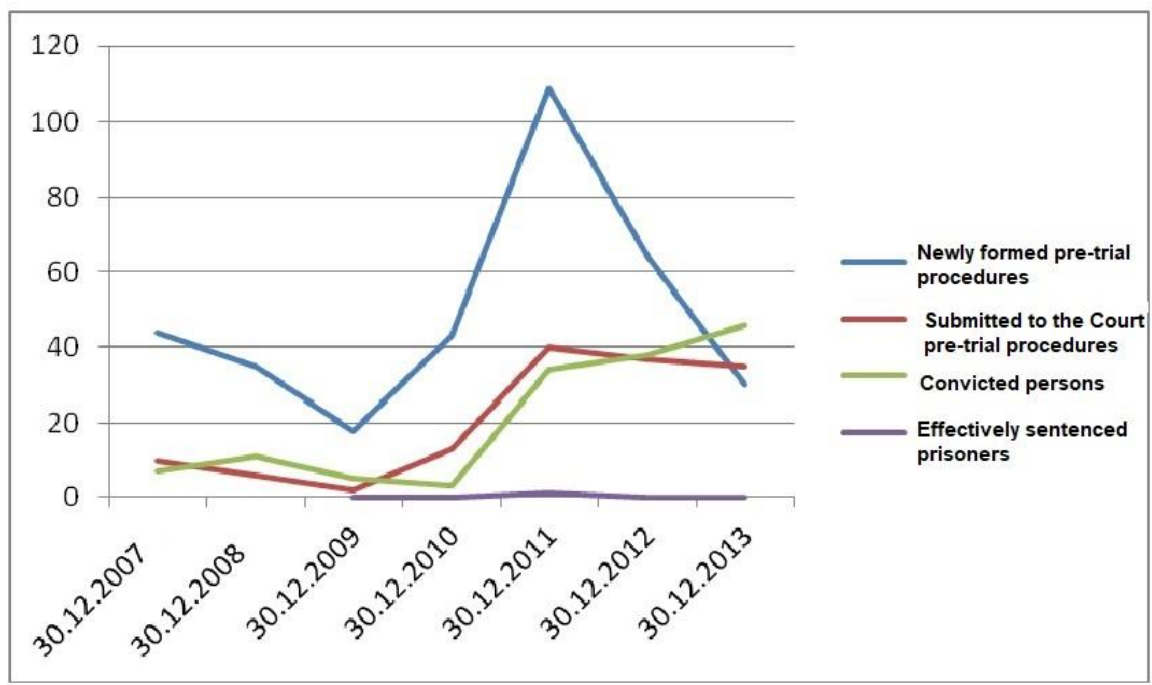

Source: author's own analysis of statistical data, provided by: Ministry of Justice, Ministry of Interior, Prosecutor's Office of Republic of Bulgaria, Ministry of Culture

Seen from the figure above it could be concluded that 25% of the registered crimes of this type by Ministry of Interior are submitted to the Court. This raises questions about the quality of the investigation, as the establishment of such type of crimes is elementary – only the illegal possession of such an object at the time of seizure is established. On the next place, there is a tendency of increase in the number of people imprisoned, as in the last years this number reaches 46 persons per year, and the average number for the entire period is approximately 14 persons per year. This shows better counteraction of the state and government, but it is ignored by the fact that only 1 person is effectively imprisoned for all committed crimes of this type for the entire country.

This indicates ($p$)=1%.

The fourth type of crimes are those related of illegal export and illegal sell or acquisition of cultural values (art. 278a of the Criminal Code). These crimes are approximately 15% of the total number of officially registered crimes of this type.

*This study was published for the first time in 2015 only in Bulgarian. The current version is its first English translation.*

Fig.6: Level of detention – crimes related to illegal export and illegal sell or acquisition of cultural values (art.278a of the Criminal Code)

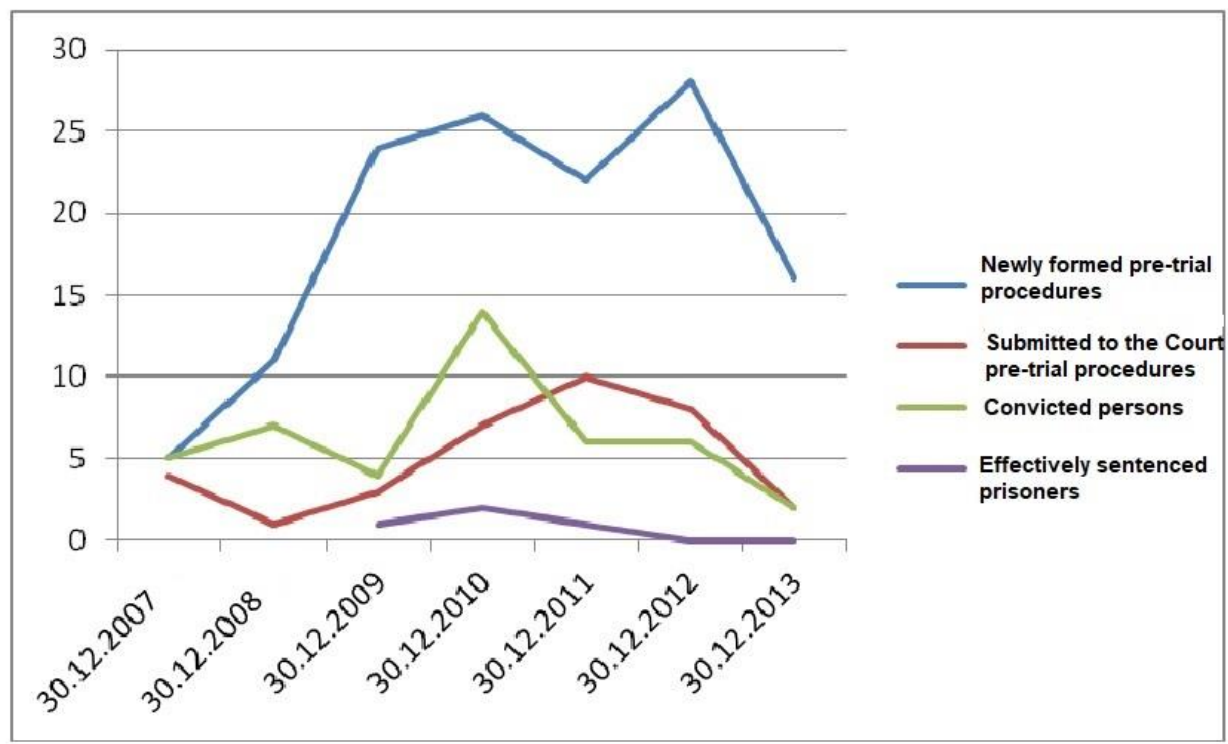

Source: author's own analysis of statistical data, provided by: Ministry of Justice, Ministry of Interior, Prosecutor's Office of Republic of Bulgaria, Ministry of Culture

Seen from the figure above, 50% of the registered crimes of this type by Ministry of Interior are submitted to the Court. This complements the low number of registered crimes of this type and at the same time, these crimes affect the top level of the criminal hierarchy of the treasure hunting. The exporters and illegal sellers of antiques are the leaders of the criminal groups. Only about 6 persons are convicted for these types of crimes per year and no one is effectively imprisoned.

This indicates $(p)=0.1\%$.

The last types of crimes are those related to the damage and destruction of cultural values (art.278b of the Criminal Code). These types of crimes are approximately 15% of the total number of officially registered crimes of this type.

*This study was published for the first time in 2015 only in Bulgarian. The current version is its first English translation.*

Fig.7: Level of detention – crimes related to the damage and destruction of cultural values (art. 278b of the Criminal Code)

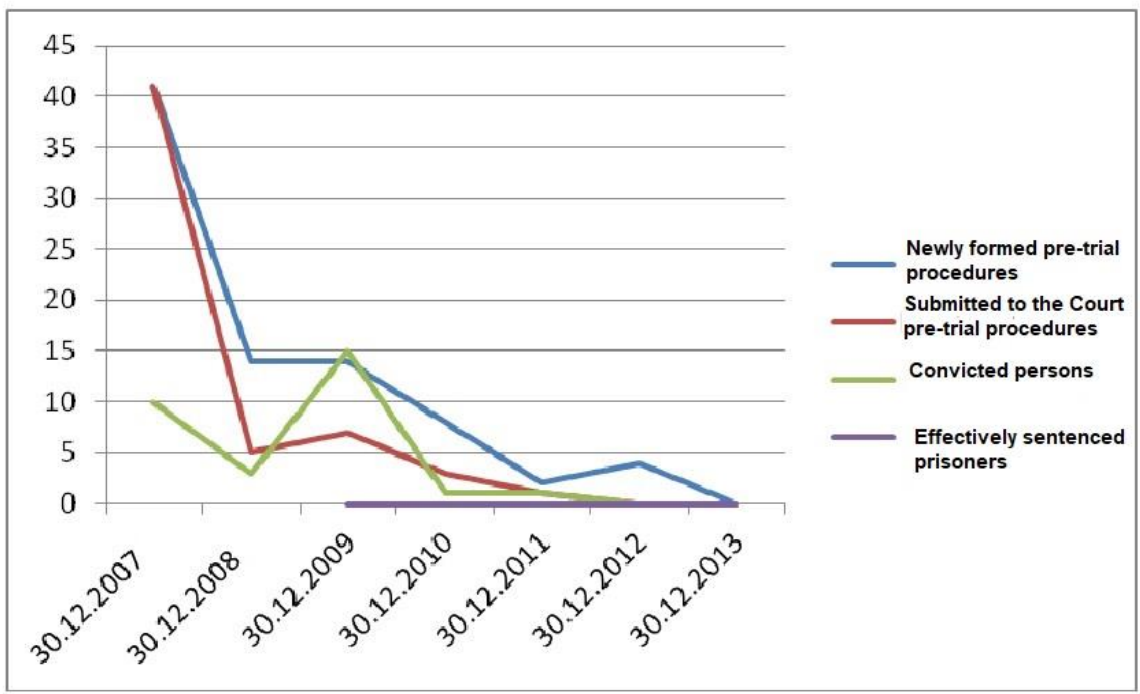

Source: author's own analysis of statistical data, provided by: Ministry of Justice, Ministry of Interior, Prosecutor's Office of Republic of Bulgaria, Ministry of Culture

The statistical data in the figure above shows that the registration, disclosure, and condemnation of criminals is close to 0. This means that the state (government) has refused to counteract to those types of crimes. Probably this due to the fact that those crimes affect the immovable cultural monuments in the construction boom period and according to the government the willingness of these people to build the centers of cities which are mainly occupied by cultural monument houses is not so publicly dangerous action. However, the public danger is related to the potential of cultural tourism development. For example, no one has ever tried to destruct the Dodge's Palace in Venice and to build a glass shopping mall on its place. These crimes cover also actions related to the destruction of Thracian tombs and sanctuaries, dug by treasure hunters looking for artefacts. In a small percentage, these types of crimes cover also the destruction of movable cultural values from different epochs, which are made of precious metal, but they do not have aesthetical value but bring scientific information which will disappear forever.

This indicates *(p)=0%.*

As a result of the empirical analysis, it could be concluded that the *risk of detention is (p)=1%.* This means that even with minimal difference between legal and illegal incomes, even those people who are risk-averse in the viewpoint of the behavioral economics will start committing crimes of this type.

**Conclusions**

Having in mind, that there is a lack of renown studies in the field of economic of crimes, examining crimes against cultural-historical and archaeological heritage, the main purpose of the current research is not to clarify all of the problems, concerning the economic analysis of these crimes.

The author of the current study has tried to modify the existing model for studying conventional crimes through the analysis of theory in the field of economics of culture and its introduction to the economics of crimes. By doing so, the author combines models that report both the direct damages caused to the individual and society by these types of crimes and the total economic value of cultural goods and values of non-use.

To a large extent, this indicates that through the opportunity costs of crimes against cultural-historical and archaeological heritage could be identified their economic dimension. However, the opportunity costs could be determined by a contingent valuation study, which is time-consuming and expensive and may be perceived as basis for future research of the author.

The author of the present study formulated a scientific thesis, which he motivated theoretically and put the basis for its future empirical and econometric confirmation.

What is important here, is that the author make difference between the categories „price"and „value", when he uses the terms optional value, existence value, educational value, prestige value donation value. The main purpose of the author is not to define a subjective category of utility – the value of cultural-historical goods, but to value them. The presented mathematical model in the first and second section of the study, even though it is a conceptual basis, shows that its development of the future would allow to be measured specific values. This is one of the main purposes of the present research.

Secondly, this study put the basis of a future analysis of crimes against cultural-historical and archaeological heritage from the viewpoint of the behavioral economics. Because of the restrictions of the study in terms of its size and volume, these types of crimes are examined mainly from the viewpoint of the government and society demand. Only in the third section – the empirical analysis, the elements of propensity to risk of the criminals and the referent point, related to the expected benefits of those risk are examined. These elements will be in depth studied by the author in a separate research, where the supply of this type of crimes, together with the equation point of demand and supply curve in the short- and long-run will be analyzed. That is why, the author has studied in depth the Freeman's model, as the author intends to develop the current research and the model of Freeman will be the hypothetic basis for considering the contributions of the author to the development of the theory and practice in this field.

On third place, the development of the current study in the future, could also include an analysis, based on discount instruments, about the damages, caused to the future generations by these types of crimes. In this way, third alternative for determining the opportunity costs will be studied: "to counteract at the maximum to those types of crimes, by redirecting the budget constraint and human capital of police, prosecution and court from other types of crimes in order to protect the cultural-historical values for future generations with the assumption in the future the cultural heritage to be a competitive advantage for Bulgaria and the country could benefit from it".

On fourth place, even though the empirical section of the research is focused on Bulgaria, this statement is not listed in the restricted conditions of research or in its topic. Because the research is the first of its kind and that is why the conclusions made could be valid for all countries. The author could develop this study in the future as he would try to test the model for other countries rich of cultural-historical and archaeological heritage in EU (Italy, Greece, Spain), neighboring countries of Bulgaria which are third countries (Turkey, North Macedonia), countries, located on other continents, for example South America, Asia. By comparison, useful guidelines for development of the relevant countries, as well as for the purposes of the international organizations such as UNESCO could be derived from such a study.

Most of the defined tasks in the introduction section of the research, such as the establishment of a conceptual framework for determining the economic dimensions of crimes against cultural-historical and archaeological heritage and their opportunity costs, are fulfilled.

In the empirical section of the research, which is tested with statistical data for Bulgaria, the author made the conclusion that the *risk of detention (p)=1%*. This means that even with minimal difference between legal and illegal incomes, even those people who are risk-averse in the viewpoint of the behavioral economics will start committing crimes against cultural-historical and archaeological heritage, which fact signals that the pre-trial procedures authorities need to increase their efforts to counteract to these types of crimes.

*This study was published for the first time in 2015 only in Bulgarian. The current version is its first English translation.*

**Notes:**

[i] e. g. Saint Cyril and Saint Metodi have been proclaimed as Patrons of Europe by Pope John Paul II

[ii] In the Constitution of the Republic of Bulgaria, it is also made a distinction amongst cultural-historic (art.23, art.54) and archaeological (art.18, p.1) heritage.

[iii] Total costs for example in the case of shoplifting, can be calculated as the average price of the stolen item is multiplied with the registered or forecasted number of shoplifting crimes. The result, however, will be sensitive mostly to the correct forecasting or registering of these types of crimes.

Average costs are based on the average price of separate type of crime and they can be used for defining in a comparative way the level of public danger of a robbery, compared to a theft.

[iv] This value is measured through contingent valuation model for the first time in USA. It is applied to judicial cases, related to ecologic damages against the American ministry of Interior at the state of Ohio, 880 F.2d 432 (US App. DC 109 1989). This valuation has been used in the court assessment of the damages, caused by the Exxon Valdez oil split, which damages are evaluated at several billion state dollars. For the first time, the idea of this value is presented by Krutilla, John, „*Conservation Reconsidered*",The American Economic Review, Volume 57, Issue 4, Sep. 1967, pp. 777-786;

[v] It was found in 1972 at the region of „Varna Chalcolithic Necropolis" dating back to the time of V Century B.C..

**References:**

1. Becker G.S. (1968) Crime and Punishment: an Economic Approach. In: Fielding N.G., Clarke A., Witt R. (eds) The Economic Dimensions of Crime. Palgrave Macmillan, London. https://doi.org/10.1007/978-1-349-62853-7_2

2. Brookshire, D., Eubanks, L., & Randall, A. (1983). Estimating Option Prices and Existence Values for Wildlife Resources. Land Economics, 59(1), 1-15. doi:10.2307/3145871

3. Brooks, Arthur C. (2002). Does Public Art Have 'Bequest Value'?, Maxwell School Working Paper, Syracuse University, Syracuse, NY.

4. Bulgarian National Television (2011). A shipment of 21,000 illegally exported antique coins and items that Canada returns to Bulgaria arrives tonight, Retrieved from: http://bnr.bg/radiobulgaria/post/100229817/pratkata-ot-21-000-nezakonno-izneseni-antichni-moneti-i-predmeti-koito-kanada-vryshta-na-bylgariya-pristiga-tazi-vecher . (in Bulgarian)

5. CSD (2007). Antiques business: traders, traffickers and connoisseurs, Organized crime in Bulgaria: markets and trends, Center for the study of democracy, 195-217. (in Bulgarian)

6. CSD (2003). Economics of crimes and anti-corruption reforms – discussions, Center for the study of democracy, 1-14; Retrieved from: www.csd.bg/fileSrc.php?id=10409 (in Bulgarian)

7. CSD (2002). New economics of crimes - discussions, Center for the study of democracy, 1-15. Retrieved from: www.csd.bg/fileSrc.php?id=10418, (in Bulgarian)

8. Constitution of Republic of Bulgaria. (2020). National Assembly of the Republic of Bulgaria Retrieved from: https://parliament.bg/en/const

9. Fol A. (1988). Thracian culture: spoken and silenced, Riva Publishing House, Sofia. (in Bulgarian)

10. Freeman, R. B. (1999). The economics of crime. Handbook of labor economics, 3, 3529-3571. https://doi.org/10.1016/S1573-4463(99)30043-2

11. Ganev, P. (2008). Economics of crimes, Institute for market economics, 1-4. Retrieved from: http://ime.bg/bg/articles/ikonomika-na-prestypleniqta/, (in Bulgarian)

12. Harris, B. S. (1984). Contingent valuation of water pollution control. Journal of Environmental Management, 19(3), 199-208.

13. ITCG, World Travel Monitor, Retrieved from: www.ipkinternational.com/en/home/

14. Krutilla, J. (1967). Conservation Reconsidered. The American Economic Review, 57(4), 777-786. http://www.jstor.org/stable/1815368

15. Lambert, R., Saunders, L., & Williams, T. (1992). Cultural sensitivity of the contingent valuation method. Lincoln University. Centre for Resource Management.

16. National Institute of Justice (2014). Training: counteracting crimes against cultural heritageл Retrieved from: http://www.nij.bg/Courses/Course.aspx?lang=bg-BG&pageid=430&ID=1725

17. Ordinance № N-3/3 December 2009 on the procedure for identification and maintenance of the Register for movable cultural values (Official journal of the republic of Bulgaria, vol.101/ 18.12.2009). in Bulgarian

18. Ordinance № N-2/12.01.2012 on determining the amount of remuneration of persons who have handed over objects in accordance with art. 93 of the Law on Cultural Heritage (Official journal of the republic of Bulgaria, vol.7/24.01.2012). in Bulgarian

19. Throsby, C. D. (1984). The measurement of willingness-to-pay for mixed goods. Oxford Bulletin of Economics and Statistics, 46(4), 279-289. https://doi.org/10.1111/j.1468-0084.1984.mp46004001.x

20. Throsby, C. D. (2010). The Economics of Cultural Policy, Cambridge University Press. https://doi.org/10.1017/CBO9780511845253

21. Weisbrod, B. A. (1964). Collective-consumption services of individual-consumption goods. The Quarterly Journal of Economics, 78(3), 471-477. https://doi.org/10.2307/1879478